\documentclass[preprint,12pt]{elsarticle}

\usepackage{amssymb}
\usepackage[table,xcdraw]{xcolor}
\usepackage{lscape}
\usepackage{longtable}
\newcommand{\etal}{\textit{et al. }}
\usepackage{subfig}
\usepackage{hyperref}
\newcommand{\ttvar}{\begingroup\@makeother\#\@ttvar}
\journal{Journal of Biomedical Informatics}

\begin{document}

 \begin{frontmatter}



\title{What do you need to consider when designing mobile health intervention?}


\author[inst1,inst2]{Aneta Lisowska}

\affiliation[inst1]{organization={Sano Centre for Computational Medicine},%Department and Organization
            % addressline={Czarnowiejska 36}, 
            city={Cracow},
            % postcode={30-054}, 
            % state={State One},
            country={Poland}}

\author[inst2]{Szymon Wilk}
\author[inst3]{Mor Peleg}

\affiliation[inst2]{organization={Institute of Computing Science, Poznan University of Technology},%Department and Organization
            % addressline={Piotrowo 3}, 
            city={Poznan},
            % postcode={60-965}, 
            % state={State Two},
            country={Poland}}
            
\affiliation[inst3]{organization={Department of Information Systems, University of Haifa},%Department and Organization
            % addressline={Yitzhak Rabin Complex}, 
            city={Haifa},
            country={Israel}}

\begin{abstract}
Designing theory-driven digital health interventions is a challenging task that needs support. We created a guide for the incomers in the field on how to design digital health interventions with case studies from the Cancer Better Life Experience (CAPABLE) European project. The guide explains how behaviour change theories can inform customisation and personalisation of the intervention. The proposed SATO (ideaS expAnded wiTh bciO) design workflow is based on the IDEAS (Integrate, Design, Assess, and Share) framework and is aligned with the Behaviour Change Intervention Ontology (BCIO). We provide a checklist of the activities that should be performed during intervention planning as well as app design templates which bundle together relevant behaviour change techniques. In the process of creating this guide, we found the necessity to extend the BCIO to support the scenarios of multiple clinical goals in the same application. The extension utilizes existing classes and properties where possible.
\end{abstract}


% \begin{highlights}
% \end{highlights}


\end{frontmatter}



% %% The Appendices part is started with the command \appendix;
% %% appendix sections are then done as normal sections
% \appendix

\section*{Statement of significance}
\noindent\textbf{Problem or issue}

\noindent Design of digital health behaviour change interventions is a complex process, yet there are few published methodologies that offer a theory-based solution that delves into the practical level of app design.   
% \vspace{5pt}
\newpage
\noindent\textbf{What is already known}

\noindent Murray \etal proposed the IDEAS (Integrate, Design, Assess, and Share) framework to support the development life-cycle of effective digital health interventions. 

\noindent Michie \etal developed classifications of behavioural change techniques (BCT) and the Behaviour Change Intervention Ontology (BCIO) to facilitate knowledge sharing and enable discovery and validation of links between successful behavioural change and particular behavioural techniques.

\vspace{5pt}
\noindent\textbf{What this paper adds}

\noindent We propose SATO (ideaS expAnded wiTh bciO) a digital health intervention design workflow aligned with the BCIO and the IDEAS frameworks.
We extend BCIO to support the scenarios of multiple behaviour change interventions (BCIs) addressing several clinical goals and extend IDEAS through integrating BCIO. Moreover, we provide a concrete example on how to build a goal hierarchy and choose evaluation measures at each level, to enable evaluation of the BCIs' effectiveness. We also prepare reusable design templates bundling together several behaviour change techniques and demonstrate how behaviour change theories can drive customisation and personalisation of the intervention.

\section{Introduction}
% Behavioural Change Intervention Ontology (BCIO) m-health applications

% Learning objectives:
The amount of literature on digital health interventions (DHI)\cite{fda} is vast \cite{klasnja2012healthcare, silva2015mobile}, with studies aiming to improve mental \cite{jameel2021mhealth} and physical health \cite{emberson2021effectiveness}. Interventions cover applications from different stages of the health management cycle and target populations across different age, social status and cultures. Some DHIs aim to support patients with adherence to pharmacological treatment, dose management and side effects reporting. Other DHIs aid patients with modulation of health risk behaviours such as: inactivity, poor nutrition, or substance abuse. The design of the latter is the focus of this paper.

Many of the industry-based DHIs do not incorporate \textit{“theory-based strategies known to drive changes in health behaviours or undergo systematic testing to demonstrate their effectiveness”} \cite{mummah2016ideas}. Some researchers, including our group, are trying to develop structured theory-based methodologies for supporting the development of effective DHIs. IDEAS (Integrate, Design, Assess, and Share) \cite{mummah2016ideas} is a framework and toolkit of strategies for the development of more effective digital interventions to change health behavior, integrating methods from behavioral theory, design thinking, and intervention evaluation and dissemination. Murray \etal draw from biomedical, behavioral, computing, and engineering research methods to develop an evaluation approach for DHIs \cite{murray2016evaluating}. They define a set of key research questions that should form the basis for an appraisal of a DHI, focusing on the target population and health need, and on factors impacting the likelihood of benefit from the DHI, considering causality, potential adverse effects, tailoring to participants, and comparison to alternative interventions.

Another source of inspiration is from Michie and colleagues. This group has been working for many years on creating a standardization of the behavior change domain; much research has been done in this domain and many behavior change interventions (BCI) have been developed. But evaluation of BCIs and comparison of theories, studies, and trials that incorporate behavior change is not feasible without a standard vocabulary and ontology. Such a structure that organizes the basic techniques that are implemented in BCIs is needed in order to attribute the success of behavioural change to particular behavioural techniques. Michie’s group developed classifications of behavioural change techniques (BCT)\cite{abraham2008taxonomy} and the Behaviour Change Intervention Ontology (BCIO) \cite{michie2020representation}.

This paper intends to provide a detailed actionable systematic guide for DHI development drawing from IDEAS  \cite{mummah2016ideas} and in alignment with the BCIO\cite{michie2020representation} and behaviour change techniques\cite{abraham2008taxonomy}. The  SATO (ideaS expAnded wiTh bciO) guide includes a design workflow and checklist (See Figure \ref{fig:workflow}). 
We provide examples taken from the Horizon 2020 project Cancer Patient Better Life Experience (CAPABLE, https://capable-project.eu/) \cite{parimbelli2021cancer}, where we follow a multi-stakeholder and evidence-based iterative development cycle for a DHI, in teams of informaticians, clinicians, patients, and engineers, and also utilize feasibility studies on synthetic and public datasets. The main contributions of this research are: 1) application and extension of the BCIO to multi-BCIs mHealth applications that span several clinical goals, 2) BCI design guide and checklist validated with examples from the CAPABLE project, and 3) reusable design templates bundling together several behaviour change techniques.

\section{Related Work}
We review the IDEAS framework and Behaviour Change Intervention (BCI) Ontology upon which we base our proposed  DBCI development guide.
\subsection{IDEAS}
IDEAS \cite{mummah2016ideas} provides a disciplined way to incrementally translate behavioural theories into highly relevant and practical interventions. Its ten steps are organized into a four-phase process: Integrate phase, including (1) empathize with target users, (2) specify target behaviour, (3) ground in behavioural theory; DEsign phase, including (4) ideate implementation strategies, (5) prototype potential products, (6) gather user feedback, (7) build a minimum viable product; Assess phase, including (8) pilot test to assess potential efficacy and usability, and (9) evaluation of the efficacy in an RCT; and Share phase, with (10) share intervention and findings. 

We previously extended IDEAS with an ontology \cite{veggiotti2021enhancing} that structures the target behaviour change intervention as a class derived from HL7 Fast Healthcare Interoperability Resources (FHIR) standard \cite{hl7} and demonstrated its application to a case study taken from the CAPABLE project, that uses Fogg's Tiny Habits behavioural model \cite{fogg2019tiny} to improve the sleep of cancer patients via Tai Chi. In another work \cite{peleg2018ideating}, we extended IDEAS Ideate step from Design phase by providing concrete backend architectural components and graphical user-interface designs that implement behavioural interventions. 

\subsection{Behaviour Change Intervention Ontology }
Ontology is an organization and representation of the entities in a domain according to their properties and relations to one another \cite{gruber1991role}. It provides shared and agreed upon definitions and descriptions of the important concepts in a domain of interests \cite{gruber1991role}.  Michie et al. \cite{michie2020representation} designed the Behaviour Change Intervention Ontology  (BCIO) following the principles of the Open Biological and Biomedical Ontology (OBO) Foundry (https://obofoundry.org) by extending the Basic Formal Ontology\cite{arp2015building}. There are six main classes in the BCIO. The first five are parts of a Behaviour Change Intervention (BCI) Scenario, in which a (Behaviour Change) Intervention, developed for a Context (ie, target population and setting), is exposed to the population via an Exposure, and via a Mechanism of Action yields an Outcome Behaviour, which is the intended new behaviour that should form a habit. The Scenario’s Outcome Behaviour --the sixth entity-- can be estimated in a clinical study. As an example for a BCI Scenario, consider a BCI of Tai Chi practice (Tai Chi Capsule), developed for the Context of fatigued cancer patients at home (see Table \ref{tab:bci-def}). The BCI/Capsule is exposed by the introduction of an app supporting the BCI that is advocated by the patients’ physician. The BCI’s Content implements a Behaviour Change Technique (BCT) \cite{abraham2008taxonomy} of demonstrating the behavior. Observing the video in which a person performs Tai Chi makes the target Outcome Behaviour clear and increases the patient's ability to perform it (hence this is the Mechanism of Action). In a BCI Evaluation Study of a BCI scenario (see the end of Table \ref{tab:bci-def}), we use an Evaluation Finding to measure the BCI Scenario’s Outcome Behaviour as well as the Intervention outcome. An Intervention outcome is a process that is influenced by an intervention, including mental activity and physiological activity (eg., sleep, stress, fatigue) and Outcome Behaviour is an Individual Human Behaviour (eg, performing Tai Chi). The Evaluation Finding objects used to measure these outcomes include a decrease in the reported Fatigue Severity Scale (FSS), a log of the number of times that a patient watched and rated Tai Chi videos, and the stress level and the number of hours slept at night, as detected from a Smartwatch. The evaluation findings at the level of BCI scenario, such as FSS, change slowly -- therefore we treat them as lag measures and check rather infrequently.
FSS measures/estimates\footnote{We use the terms "measure" and "estimate" interchangeably; "estimate" to be consistent with BCIO \cite{michie2020representation} and "measure" to be aligned with the 4DX terminology \cite{mcchesney20124}} a patient's fatigue (current state). The change in the fatigue rating is the \textbf{\textit{intervention outcome estimate/lag measure}}.
The number of times a patient played a Tai Chi video reflects the process of engaging with the intervention and could be a \textbf{\textit{lead measure}} for intervention outcome. Lead measures track the critical activities that drive or lead to the \textbf{\textit{lag (outcome) measure}} \cite{mcchesney20124}.

% We extent the BCI
\begin{landscape}
% \caption{BCIO examples for the Fatigue Improvement Scenario

\begin{longtable}{|l|l|} 
\caption{BCIO examples for the Fatigue Reduction Scenario}\\ 
\hline
{\color[HTML]{3531FF}Class} with {\color[HTML]{F076B5}properties}                        & {\color[HTML]{009901} Individual}                                   \\ 
\hline
{\color[HTML]{3531FF}BCI\_Scenario}& {\color[HTML]{009901} Fatigue Reduction Scenario}                  \\
{\hspace{5pt}\color[HTML]{F076B5}has\_BCI\_context} {\color[HTML]{3531FF}{[}BCI\_Context{]} } & {\hspace{5pt}\ \color[HTML]{009901} Cancer patient with fatigue problems at home} \\
{\hspace{5pt}\color[HTML]{F076B5}has\_occurant\_part} {\color[HTML]{3531FF}{[}BCI{]}} & {\color[HTML]{009901} \begin{tabular}[c]{@{}l@{}}\hspace{5pt} \{Tai\_Chi\_Capsule, Walk\_Capsule, \\ \hspace{5pt}Deep\_Breathing\_Capsule,  Imagery\_Training\_Capsule\}\end{tabular}} \\
\begin{tabular}[c]{@{}l@{}}{\textbf{\hspace{5pt}\color[HTML]{F076B5}*has\_goal} {\color[HTML]{3531FF}{[}Clinical\_Goal - information content entity{]}}}\end{tabular} &
  {\color[HTML]{009901} \begin{tabular}[c]{@{}l@{}} \hspace{5pt} Less fatigue (as measured by Fatigue Severity Scale \textless{}36)\end{tabular}} \\
{\textbf{\hspace{5pt}\color[HTML]{F076B5}*has\_occurant\_part }{\color[HTML]{3531FF}{[}Intervention Outcome{]}}} & {\hspace{5pt} \color[HTML]{009901} Change in reported Fatigue Severity Scale}    \\ 
\hline
{\color[HTML]{3531FF}BCI}& {\color[HTML]{009901} Tai\_Chi\_Capsule}                  \\
{\hspace{5pt}\color[HTML]{F076B5}has\_occurent\_part} {\color[HTML]{3531FF}{[}BCI\_Content{]} } & {\hspace{5pt}\ \color[HTML]{009901} Tai\_Chi\_Capsule\_Content} \\
{\hspace{5pt}\color[HTML]{F076B5}has\_occurant\_part} {\color[HTML]{3531FF}{[}BCI\_Delivery{]}} & {\color[HTML]{009901} \begin{tabular}[c]{@{}l@{}}\hspace{5pt} Tai\_Chi\_Delivery \end{tabular}} \\
\begin{tabular}[c]{@{}l@{}}{\textbf{\hspace{5pt}\color[HTML]{F076B5}*has\_BCI\_Source} {\color[HTML]{3531FF}{[}BCI\_Source{]}}}\end{tabular} &
  {\color[HTML]{009901} \begin{tabular}[c]{@{}l@{}} \hspace{5pt} Source: Song \etal \cite{song2018ameliorative}\end{tabular}} \\
{\textbf{\hspace{5pt}\color[HTML]{F076B5}*has\_occurant\_part }{\color[HTML]{3531FF}{[}Outcome Behaviour{]}}} & {\hspace{5pt} \color[HTML]{009901} Performing Tai Chi regularly}    \\ 
{\textbf{\hspace{5pt}\color[HTML]{F076B5}*has\_occurant\_part }{\color[HTML]{3531FF}{[}BCI\_Mechanism\_of\_Action{]}}} & { {\color[HTML]{009901}\begin{tabular}[c]{@{}l@{}}\hspace{5pt}  Performing Tai Chi habitually relaxes the mind and body \\\hspace{5pt} and allows the person   to fall asleep more rapidly and to   \\\hspace{5pt} stay sleeping for longer stretches of time\end{tabular} }} \\ 
\hline
\rowcolor[HTML]{C0C0C0} a) &\\
{\color[HTML]{3531FF}BCI\_Content}& {\color[HTML]{009901} Habit\_planning\_for\_Tai\_Chi\_Content}                  \\
{\hspace{5pt}\color[HTML]{F076B5}has\_profile} {\color[HTML]{3531FF}{[}BCI\_Dose{]} } & {\hspace{5pt}\ \color[HTML]{009901}  1/week (review)} \\
{\hspace{5pt}\color[HTML]{F076B5}has\_occurant\_part} {\color[HTML]{3531FF}{[}BC\_Technique{]}} & {\color[HTML]{009901} \begin{tabular}[c]{@{}l@{}}\hspace{5pt}\{[BCT4];[BCT10]; [BCT15]; [BCT26]\} \end{tabular}} \\
\hspace{10pt} {\color[HTML]{3531FF}{BC\_Technique}} & {\color[HTML]{009901} \begin{tabular}[c]{@{}l@{}}\hspace{10pt}[BCT4]Prompt intention formation
\end{tabular}}  \\
\begin{tabular}[c]{@{}l@{}}{\textbf{\hspace{15pt}\color[HTML]{F076B5}*has\_occurent\_part} {\color[HTML]{3531FF}{[}BCI\_Mechanism\_of\_Action{]}}}\end{tabular} &
  {\color[HTML]{009901} \begin{tabular}[c]{@{}l@{}} \hspace{15pt} Behavioural resolution reinforces goal acceptance \end{tabular}} \\
\hspace{10pt} {\color[HTML]{3531FF}{BC\_Technique}} & {\color[HTML]{009901} \begin{tabular}[c]{@{}l@{}}\hspace{10pt}[BCT10]Prompt specific goal setting
 \end{tabular}} \\
\begin{tabular}[c]{@{}l@{}}{\textbf{\hspace{15pt}\color[HTML]{F076B5}*has\_occurent\_part} {\color[HTML]{3531FF}{[}BCI\_Mechanism\_of\_Action{]}}}\end{tabular} &
  {\color[HTML]{009901} \begin{tabular}[c]{@{}l@{}} \hspace{15pt} Specification of at least one context, that is, where, \\\hspace{15pt} when,  how makes the brain establish a "memory" \\ \hspace{15pt} that makes the practice easier  \end{tabular}} \\
\hspace{10pt} {\color[HTML]{3531FF}{BC\_Technique}} & {\color[HTML]{009901} \begin{tabular}[c]{@{}l@{}}\hspace{10pt}[BCT15]Teach to use prompts or cues
 \end{tabular}} \\
\begin{tabular}[c]{@{}l@{}}{\textbf{\hspace{15pt}\color[HTML]{F076B5}*has\_occurent\_part} {\color[HTML]{3531FF}{[}BCI\_Mechanism\_of\_Action{]}}}\end{tabular} &
  {\color[HTML]{009901} \begin{tabular}[c]{@{}l@{}} \hspace{15pt}  Learning to identify a prompt and react to it makes \\\hspace{15pt} the  response more instinctive  \end{tabular}} \\ 
\hspace{10pt} {\color[HTML]{3531FF}{BC\_Technique}} & {\color[HTML]{009901} \begin{tabular}[c]{@{}l@{}}\hspace{10pt}[BCT26]Time management
 \end{tabular}} \\
\begin{tabular}[c]{@{}l@{}}{\textbf{\hspace{15pt}\color[HTML]{F076B5}*has\_occurent\_part} {\color[HTML]{3531FF}{[}BCI\_Mechanism\_of\_Action{]}}}\end{tabular} &
  {\color[HTML]{009901} \begin{tabular}[c]{@{}l@{}} \hspace{15pt} Freeing time for the behaviour makes it possible  \\\hspace{15pt} to practice it \end{tabular}} \\ 
%  \\
{\textbf{\hspace{4pt}\color[HTML]{F076B5}*has\_occurant\_part }{\color[HTML]{3531FF}{[}BCI\_Engagement{]}}} & {\hspace{5pt} \color[HTML]{009901}  \{set the goal, the prompt and the time for the activity\}}
   \\ 
{\textbf{\hspace{4pt}\color[HTML]{F076B5}*has\_occurant\_part }{\color[HTML]{3531FF}{[}BCI\_Delivery{]}}} & { {\color[HTML]{009901}\begin{tabular}[c]{@{}l@{}}\hspace{5pt} Habit\_planning\_for\_Tai\_Chi\_Delivery \end{tabular} }} \\ 

\hspace{10pt} {\color[HTML]{3531FF}{BCI\_Delivery}} & {\color[HTML]{009901} \begin{tabular}[c]{@{}l@{}}\hspace{10pt}
 \end{tabular}} \\
\begin{tabular}[c]{@{}l@{}}{\hspace{15pt}\color[HTML]{F076B5}has\_BCI\_Source} {\color[HTML]{3531FF}{[}BCI\_Source{]}}\end{tabular} &
  {\color[HTML]{009901} \begin{tabular}[c]{@{}l@{}} \hspace{15pt}   \end{tabular}} \\ 
\begin{tabular}[c]{@{}l@{}}{\hspace{15pt}\color[HTML]{F076B5}has\_profile} {\color[HTML]{3531FF}{[}BCI\_Mode\_of\_delivery{]}}\end{tabular} &
  {\color[HTML]{009901} \begin{tabular}[c]{@{}l@{}} \hspace{15pt}  mobile\_app\_mode\_of\_delivery \end{tabular}} \\
  \begin{tabular}[c]{@{}l@{}}{\hspace{15pt}\color[HTML]{F076B5}has\_profile} {\color[HTML]{3531FF}{[}BCI\_Schedule\_of\_delivery{]}}\end{tabular} &
  {\color[HTML]{009901} \begin{tabular}[c]{@{}l@{}} \hspace{15pt}  @onboarding \end{tabular}} \\
  \begin{tabular}[c]{@{}l@{}}{\hspace{15pt}\color[HTML]{F076B5}has\_profile} {\color[HTML]{3531FF}{[}BCI\_Style\_of\_delivery{]}}\end{tabular} &
  {\color[HTML]{009901} \begin{tabular}[c]{@{}l@{}} \hspace{15pt}   \end{tabular}} \\
  \rowcolor[HTML]{C0C0C0} b) &\\
{\color[HTML]{3531FF}BCI\_Content}& {\color[HTML]{009901} Tai\_Chi\_Lesson\_Content}                  \\
{\hspace{5pt}\color[HTML]{F076B5}has\_profile} {\color[HTML]{3531FF}{[}BCI\_Dose{]} } & {\hspace{5pt}\ \color[HTML]{009901}  1/day} \\
{\hspace{5pt}\color[HTML]{F076B5}has\_occurant\_part} {\color[HTML]{3531FF}{[}BC\_Technique{]}} & {\color[HTML]{009901} \begin{tabular}[c]{@{}l@{}}\hspace{5pt}\{[BCT9], [BCT12]\} \end{tabular}} \\
\hspace{10pt} {\color[HTML]{3531FF}{BC\_Technique}} & {\color[HTML]{009901} \begin{tabular}[c]{@{}l@{}}\hspace{10pt}[BCT9]Model or demonstrate the behaviour
 \end{tabular}} \\
\begin{tabular}[c]{@{}l@{}}{\textbf{\hspace{15pt}\color[HTML]{F076B5}*has\_occurent\_part} {\color[HTML]{3531FF}{[}BCI\_Mechanism\_of\_Action{]}}}\end{tabular} &
  {\color[HTML]{009901} \begin{tabular}[c]{@{}l@{}} \hspace{15pt} Observing a person perform the activity makes it  \\\hspace{15pt} clear and increases patient's ability to perform it  \end{tabular}} \\
 \hspace{10pt} {\color[HTML]{3531FF}{BC\_Technique}} & {\color[HTML]{009901} \begin{tabular}[c]{@{}l@{}}\hspace{10pt}[BCT12] Prompt self-monitoring of behaviour.
 \end{tabular}} \\
\begin{tabular}[c]{@{}l@{}}{\textbf{\hspace{15pt}\color[HTML]{F076B5}*has\_occurent\_part} {\color[HTML]{3531FF}{[}BCI\_Mechanism\_of\_Action{]}}}\end{tabular} &
  {\color[HTML]{009901} \begin{tabular}[c]{@{}l@{}} \hspace{15pt} The  ticking of the box after video rewards behaviour\\\hspace{15pt} and feed-backs lesson completion \end{tabular}} \\
  
{\textbf{\hspace{4pt}\color[HTML]{F076B5}*has\_occurant\_part }{\color[HTML]{3531FF}{[}BCI\_Engagement{]}}} & {\hspace{5pt} \color[HTML]{009901}  \{clicking play the video\}}
   \\ 
{\textbf{\hspace{4pt}\color[HTML]{F076B5}*has\_occurant\_part }{\color[HTML]{3531FF}{[}BCI\_Delivery{]}}} & { {\color[HTML]{009901}\begin{tabular}[c]{@{}l@{}}\hspace{5pt} Demonstrating\_Tai\_Chi\_Delivery \end{tabular} }} \\ 
\hspace{10pt} {\color[HTML]{3531FF}{BCI\_Delivery}} & {\color[HTML]{009901} \begin{tabular}[c]{@{}l@{}}\hspace{10pt}
 \end{tabular}} \\
\begin{tabular}[c]{@{}l@{}}{\hspace{15pt}\color[HTML]{F076B5}has\_BCI\_Source} {\color[HTML]{3531FF}{[}BCI\_Source{]}}\end{tabular} &
  {\color[HTML]{009901} \begin{tabular}[c]{@{}l@{}} \hspace{15pt} LeiaCohen@Taiflow  \end{tabular}} \\ 
\begin{tabular}[c]{@{}l@{}}{\hspace{15pt}\color[HTML]{F076B5}has\_profile} {\color[HTML]{3531FF}{[}BCI\_Mode\_of\_delivery{]}}\end{tabular} &
  {\color[HTML]{009901} \begin{tabular}[c]{@{}l@{}} \hspace{15pt}  playable\_electronic\_storage\_mode\_of\_delivery \end{tabular}} \\
  \begin{tabular}[c]{@{}l@{}}{\hspace{15pt}\color[HTML]{F076B5}has\_profile} {\color[HTML]{3531FF}{[}BCI\_Schedule\_of\_delivery{]}}\end{tabular} &
  {\color[HTML]{009901} \begin{tabular}[c]{@{}l@{}} \hspace{15pt} preferred\_time\_of\_day  \end{tabular}} \\
  \begin{tabular}[c]{@{}l@{}}{\hspace{15pt}\color[HTML]{F076B5}has\_profile} {\color[HTML]{3531FF}{[}BCI\_Style\_of\_delivery{]}}\end{tabular} &
  {\color[HTML]{009901} \begin{tabular}[c]{@{}l@{}} \hspace{15pt} warm\_and\_accepting  \end{tabular}} \\
 \hline
 {\color[HTML]{3531FF}Intervention\_Evaluation\_Study}& {\color[HTML]{009901} Fatigue Reduction Scenario Intervention Study}                  \\
{\hspace{5pt}\color[HTML]{F076B5} evaluates} {\color[HTML]{3531FF}{[}BCI\_Scenario{]} } & {\hspace{5pt}\ \color[HTML]{009901} Fatigue Reduction Intervention Scenario } \\
{\hspace{5pt}\color[HTML]{F076B5}has\_output} {\color[HTML]{3531FF}{[}Evaluation\_Finding{]}} & \\
{ \hspace{10pt}\color[HTML]{3531FF}{ Evaluation\_Finding}} &
{\color[HTML]{009901} \begin{tabular}[c]{@{}l@{}}\hspace{5pt} Decreased Fatigue as measured by Severity Scale-Lag\end{tabular}} \\
% \\
\hline
 {\color[HTML]{3531FF}BCI\_Evaluation\_Study}& {\color[HTML]{009901} Tai Chi Intervention Study}                  \\
{\hspace{5pt}\color[HTML]{F076B5} evaluates} {\color[HTML]{3531FF}{[}BCI{]} } & {\hspace{5pt}\ \color[HTML]{009901}Tai\_Chi\_Capsule } \\
{\hspace{5pt}\color[HTML]{F076B5}has\_output} {\color[HTML]{3531FF}{[}Evaluation\_Finding{]}} & \\
{ \hspace{10pt}\color[HTML]{3531FF}{Evaluation\_Finding}} &
{\color[HTML]{009901} \begin{tabular}[c]{@{}l@{}}\hspace{9pt}  \{Num\_times\_Tai\_chi\_Video\_Pressed-Lead; \\\hspace{10pt}  Stress\_decreased\_Smartwatch;  \\\hspace{10pt}  Num\_hours\_slept\_a\_night\_increased\_Smartwatch\}
\end{tabular}} 

\label{tab:bci-def}\\
\hline
\end{longtable}

\end{landscape}

\section{Digital Intervention Design Steps}
\begin{figure}
    \centering
    \includegraphics[width=1.0\linewidth]{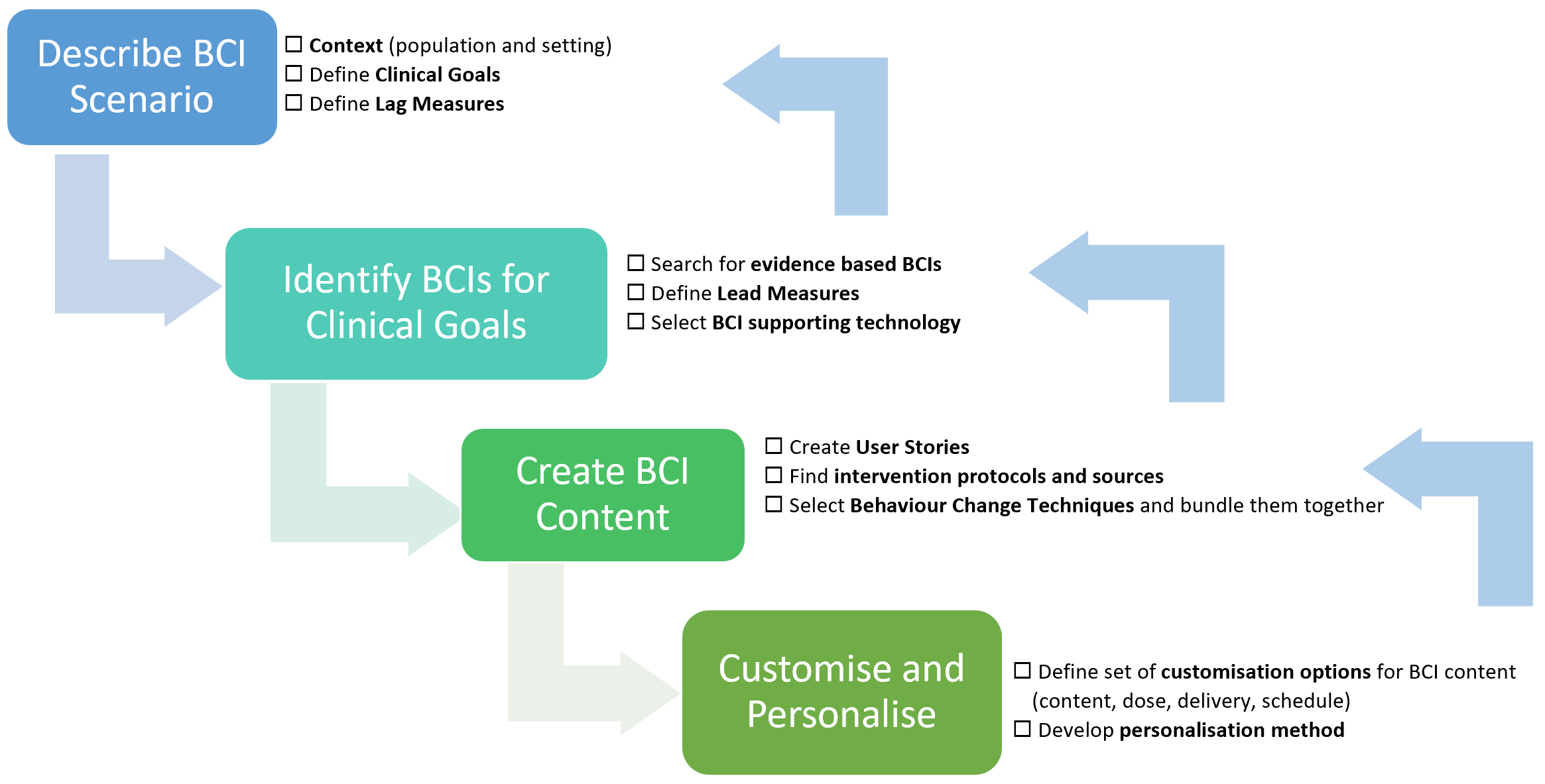}
    \caption{SATO workflow for development of DHI apps that create desirable behaviour change. }
    \label{fig:workflow}
\end{figure}

An important contribution of this study is to suggest a  SATO (ideaS expAnded wiTh bciO) workflow for the development of DHI apps that create desirable behaviour change. We consider DHI that corresponds to BCI Scenario and is a collection of BCIs. Figure \ref{fig:workflow} summarizes the workflow and the sections below elaborate on the different activities of the workflow. 

\begin{figure} [ht]
    \centering
    \includegraphics[width=1.0\linewidth]{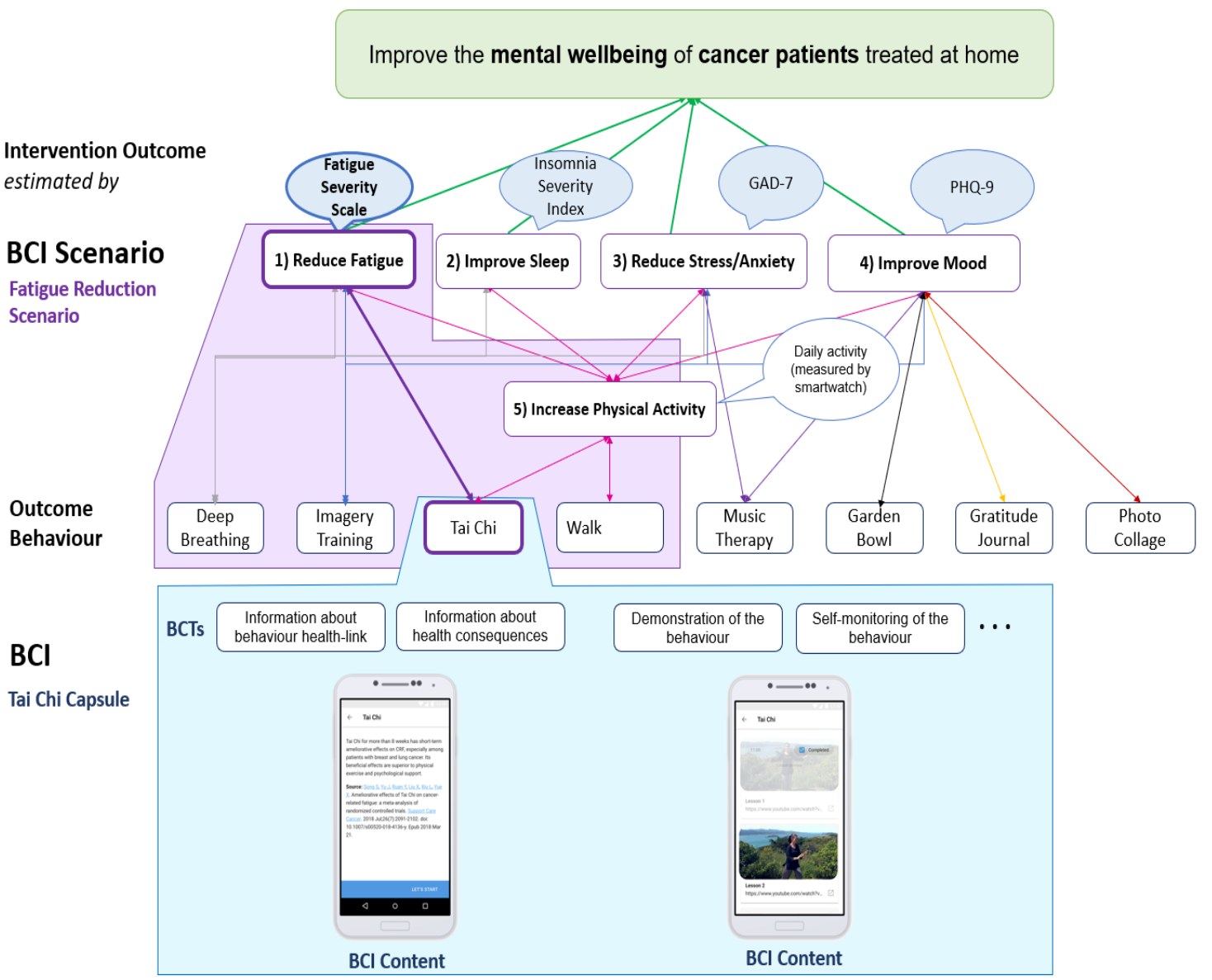}
    \caption{BCI Scenario, BCI, and BCT examples from the CAPABLE project}
    \label{fig:bci_scenario}
\end{figure}

We focus the SATO workflow on methodological construction of a Behaviour Change Intervention Scenario Plan for the Behaviour Change Intervention Scenario process. We highlight our extensions of the BCIO based on our need to define clinical goals for the BCI Context (population and setting) – and for each goal -– to recommend, specify and develop BCIs meeting the goal. Based on our experience of implementing several BCIs as part of the CAPABLE project, our running example is a BCI Scenario in which the Behaviour Outcome is practice of Tai Chi and the desired Intervention Outcome is a reduction of fatigue, as measured by the standard Fatigue Severity Scale  \cite{krupp1989fatigue} (Figure \ref{fig:bci_scenario}). 

\subsection{BCI Scenario}

Following the IDEAS framework for developing effective digital interventions to change health behaviour \cite{mummah2016ideas}, we recommend that the DHI will be developed by a multi-stakeholder development team consisting of the clinicians whose patients will use the DHI, the patients themselves, and the technical developers (engineers, informaticians), which would meet regularly during the DHI development phase.

\subsubsection{Understand the Context}
% \subsubsection{Population}
The first step that this team does is defining the BCI Scenario’s patient \textbf{population}. For the CAPABLE project, it is cancer patients treated with immunotherapy. For the fatigue improvement intervention scenario, we focus on the subset of patients who experience fatigue.

The intervention \textbf{setting} (eg, patients treated at home) determines the possible mode of delivery of the BCI and influences the barriers to engagement. Considering that the patients are treated at home imposes a constraint on the delivery mode, which is limited to digital interventions. Potential barriers to engagement with DHI in this context could be a patient’s lack of access or lack of ability to use the digital technology (such as wearable devices, eg smartwatches).

\subsubsection{Identify Clinical Goal (extension of BCIO)}
A crucial step in developing a DHI is to select the important clinical objective or goal of the DHI. The clinical goal will impact the choice of intervention and evaluation metrics.

The clinical goal can be applicable to a diverse group of patients because it may be a condition or avoiding the adverse effect that is typical for many kinds of patients, like reducing sleep problems, anxiety, or increasing physical activity. However, prior definition of a more specific context, ie, the target population (eg, cancer patients) and the physical and social setting for the behaviour change, may help in the creation of a targeted DHI that can address the wellbeing needs specific to the population, recognize their disease burden and support personalization of the interventions. Therefore, to select the clinical goals of the DHI for the target population, research should be done to establish what wellbeing dimensions \cite{linton2016review} are the ones most impacted by the patients’ condition,  fitting with IDEAS' Empathize step of the Integrate phase, that integrates insights from users and theory. This step should be led by the clinical researchers via a literature search supplemented by questionnaires to the target population. 
For example, in the CAPABLE project, our goal is to \textbf{improve the mental wellbeing of cancer patients treated at home}. The main affected wellbeing dimensions were identified from the literature and from questionnaires answered by renal cancer patients in Pavia, Italy, melanoma cancer patients in Amsterdam, The Netherlands, and patients from the Italian Cancer Patient organization, AIMAC. They include reduced sleep quality \cite{mystakidou2007sleep}, fatigue \cite{hofman2007cancer}, stress \cite{zabora2001prevalence}, low mood \cite{spiegel2003depression} and decreased physical activity \cite{littman2010longitudinal}. Improvements in all these dimensions are then posed as clinical goals (See top of Figure \ref{fig:bci_scenario}). The clinical goals might be intertwined, (e.g. increasing physical activity could be a goal in itself but also benefits sleep quality) and should be targeted by the interventions provided by the digital health app. 

\begin{figure} [ht]
    \centering
    \includegraphics[width=1.0\linewidth]{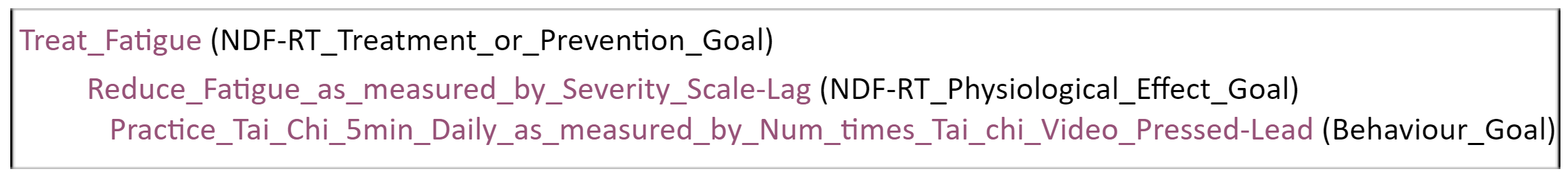}
    \caption{Goal hierarchy for the Fatigue Treatment Goal}
    \label{fig:goal_hierarchy}
\end{figure}

\textbf{Goal ontologies} can be used to standardize the specification of clinical goals. For example, in the Goal-based Comorbidity decision-support method \cite{kogan2020towards}, goals specification follows the goal ontology developed by Fox et al. \cite{fox2006ontological} that includes a verb and a noun phrase (eg, manage hypertension, prevent cardiovascular disease, treat fatigue), the HL7 Fast Healthcare Interoperability Resources (FHIR) \cite{hl7} Goal resource, and relationships from the National Drug File - Reference Terminology (NDF-RT) ontology \cite{ndf}, such as physiological effects of medications (eg, decrease platelet-aggregation, increased sleep). Figure \ref{fig:goal_hierarchy} shows an example of such a goal hierarchy that goes all the way down to a leaf BCI meeting the goal. Alternatively, in the Asbru \cite{shahar1997task} clinical guideline formalism, process goals (e.g., monitor blood pressure) or state goals (e.g., normal blood pressure) can be specified as temporal patterns that are meant to be maintained, avoided or achieved (e.g., achieve systolic blood pressure $<$140 within 1 month of starting antihypertensive medication). In our fatigue reduction intervention scenario, "no clinically significant fatigue" would be a state goal and daily performance of Tai Chi a process (behaviour) goal.

\subsubsection{Define Lag Measures}

How do we measure the intervention's effectiveness? According to Michie et al \cite{michie2020representation}, intervention outcomes are processes influenced by interventions which include individual human behaviour, mental activity and physiological activity, including unintended side effects. They are measured as BCI evaluation finding, including BCI effect estimate and BCI outcome estimate. During the enrollment and at the termination of the CAPABLE study, patients are asked to fill multiple standard patient-reported outcome measures (PROMs) \cite{cella2015patient}, such as Insomnia Severity Index (ISI) \cite{morin2011insomnia}, Fatigue Severity Scale (FSS) \cite{krupp1989fatigue}, GAD-7 \cite{williams2014gad} and PHQ-9 \cite{kroenke2001phq}, which enable assessment of patients' mental health prior and post intervention. Each of these questionnaires is linked with a clinical goal. Changes in scores on these questionnaires do not occur rapidly; therefore the changes in scores between enrolment and exit questionnaires are our lag measures\cite{mcchesney20124}. For example, for the goal of improved fatigue, the patients might score above 36 on the FSS at the enrollment, which indicates clinical fatigue. Ideally after a few months, at the termination of the intervention, the patient would have FSS less than 36, corresponding to no clinically significant fatigue. Note that for the BCI Scenario we measure the change in patients’ physiological/emotional state related to their clinical goal and the performance of the target behaviours is considered at the BCI level (See Table \ref{tab:bci-def}).

\subsection{Identify BCIs for the clinical goals}

To refine the clinical goals and identify the interventions that can meet the goals, we can turn to clinical practice guidelines. Following the evidence-based medicine (EBM) movement, “clinical practice guidelines that we can trust” are defined as “statements that include recommendations intended to optimize patient care, that are informed by a systematic review of evidence and an assessment of the benefit and harms of alternative care options for a clinical condition” (clinical objective/goal) \cite{graham2011institute}. \textbf{Clinical guidelines} usually address a specific clinical condition. Unfortunately, most of the clinical guidelines refer to medication-based care options, and evidence for non-medication interventions is usually limited. However, non-pharmacological life-style, exercise, and psycho-behavioural interventions are a promising way to care for mental wellbeing, including for example, chronic pain\cite{chou2007nonpharmacologic} and fatigue \cite{fabi2020cancer}, and include evidence grades \cite{muradnew} based on cohort studies, and in some cases on randomized controlled trials and meta-analyses, which provide a higher grade of evidence. 

\subsubsection{Search for evidence-based intervention}
Clinical goal(s) is an important extension of the BCIO and searching for evidence-based intervention options that meet it fits the “specify target behaviour” step of the Integrate phase of IDEAS. As mentioned above, clinical practice guidelines, and other clinical sources following the EBM pyramid, are the best way to search for evidence-based interventions. For example, the ESMO cancer-related fatigue guideline \cite{fabi2020cancer} suggests mindfulness-based stress reduction techniques including deep breathing, imagery training, Hatha yoga, walking in nature, and physical exercise. Additional evidence-based meta-reviews \cite{muradnew} suggests Tai Chi \cite{song2018ameliorative} as a further option. Figure \ref{fig:bci_scenario} shows interventions that meet the different clinical goals. We also added the source property, which originally is a property of a BCT, to BCI to specify its evidence-based source (See Table \ref{tab:bci-def}).

\subsubsection{Define Lead Measures}

To ensure that the BCI is effective in reaching the target clinical goal, ideally we would be able to check if a patient is on track of reaching their goal, and if not, modify the intervention. However, daily assessment through patient-reported outcome measure (PROM) questionnaires is not feasible long term, especially given that previous studies found that frequent surveys were not perceived favorable by the study participants \cite{amorim2019integrating} and could negatively impact engagement with the intervention. Therefore it is important to identify measures that are related to the outcome but can also be captured frequently and automatically. In order to reduce the patient burden of self-reporting, we opted out of daily wellbeing surveys and settled on tracking proxies of wellbeing, such as physical activity (e.g., number of steps) and quality and duration of sleep and stress that can be passively measured by a smartwatch \cite{cornet2018systematic}. In our pilot study on a healthy population, we found that the number of times participants went for a walk correlated with the magnitude of fatigue reduction \cite{lisowska2022pilot}. Given that each BCI has a target behaviour, whose performance should support patients in reaching their clinical goal, the number of times that a target behaviour has been performed could be a good lead measure. For example, the number of times a patient performs Tai Chi may serve as a lead measure for reduction of fatigue. For some goals (e.g., sleep improvement as measured by Insomnia Severity Index) there could also be automatic daily measurement of number of hours slept a night, which may indicate early if the intervention leads towards the achievement of the clinical goal.

\subsubsection{Select BCI supporting technologies}

To monitor intervention adherence, it might be helpful to pair the DHI app with a wearable device. The choice of the device will depend on the population, their clinical goals, the selected BCIs and the lead measures. In the CAPABLE project, the wearable device (ASUS VivoWatch) was selected by the clinical team based on the feature of automatic measurement of Pulse Transit Time index \cite{ghosh2016continuous} and accessibility of the data by a web-based API.

Detecting performance of the target behaviours using wearable devices could be straightforward for some BCIs (e.g., walk); for other behaviours,  e.g., Tai Chi or Deep Breathing exercises, detection might be more challenging. Consumer-grade smartwatches are frequently equipped with a photoplethysmogram (PPG) sensor that could capture blood volume pulse (BVP) signal \cite{saganowski2020review}. Therefore, based on experiments on a public dataset and a literature review, we hypothesized that we could detect when a patient performed deep breathing exercise from the changes to their BVP \cite{lisowska2021catching}. Nevertheless, during initial evaluation of the device we found that the ASUS API exposes only extracted heart rate variability features rather than the raw PPG signal. The feature is captured in 15 minutes time steps, which results in its limited usefulness for assessment of adherence to 5 minute-long deep breathing exercises. When selecting a wearable device we suggest to consider not only \textbf{type} of captured data but also the \textbf{frequency} and to test the candidate devices early in the application development cycle.

For deep breathing and Tai Chi, an easier step for adherence monitoring is to record mobile application’s usage data, including information such as if and when the patient started the exercise recording embedded within the application. This captures engagement with the application; not necessarily the performance of the behaviour \cite{yardley2016understanding}. Yet it is a simple solution, not requiring a specialized wearable device and could serve as a BCI adherence proxy without posing additional reporting burden on the patient.

\subsection{Create BCI Content}
\subsubsection{Create user stories}
\begin{figure} [ht]
    \centering
    \includegraphics[width=1.0\linewidth]{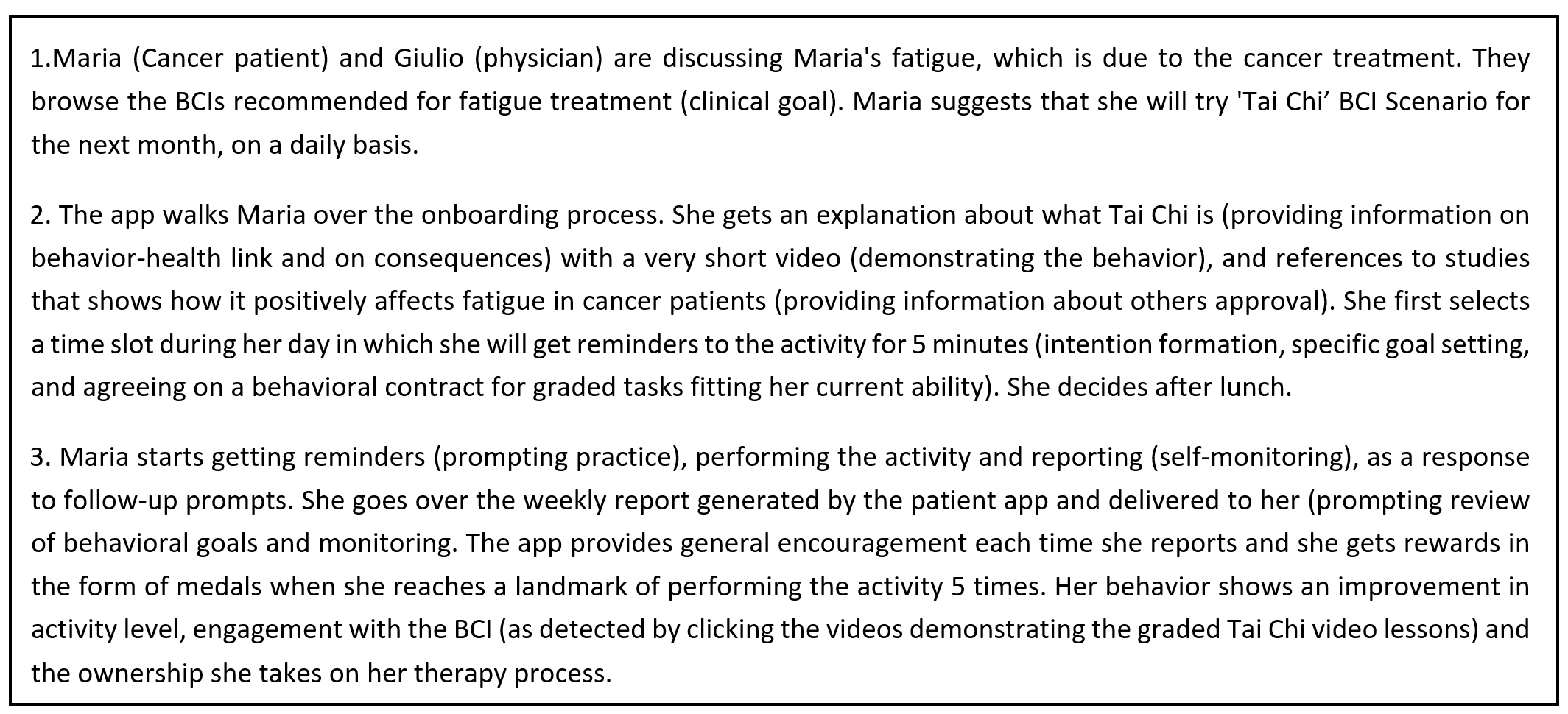}
    \caption{A set of user stories for a patient persona called Maria and her physician Giulio involving the Tai Chi BCI for the fatigue goal and following BCTs.
}
    \label{fig:user_story}
\end{figure}

\begin{figure}%
    \centering
    \subfloat[\centering Goal-setting in physicians' app]{{\includegraphics[width=10cm]{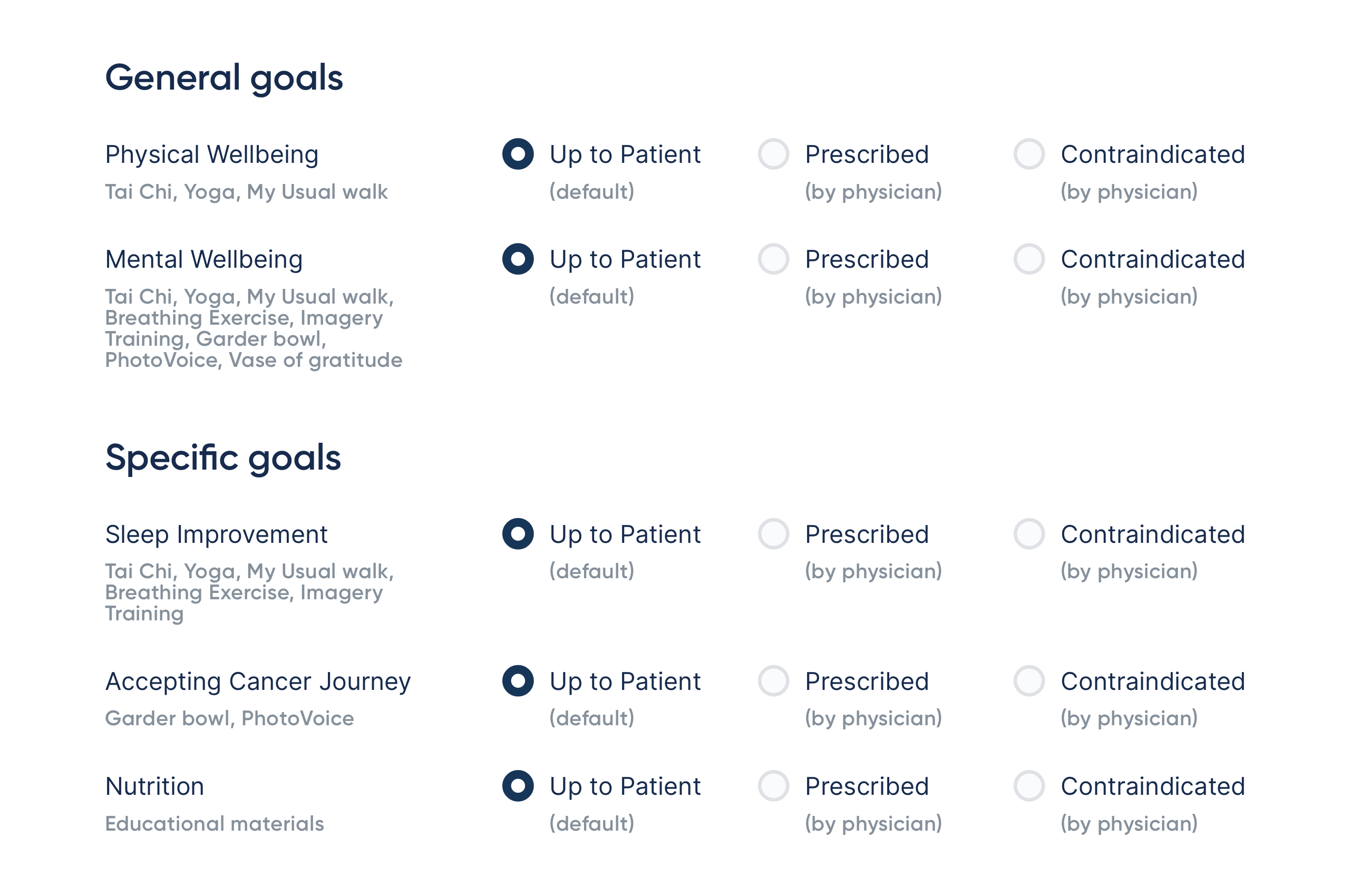}
        \label{mphys}
    }}%
    \\
        \subfloat[\centering Goals screen in patient app ]{{\includegraphics[width=5.0cm]{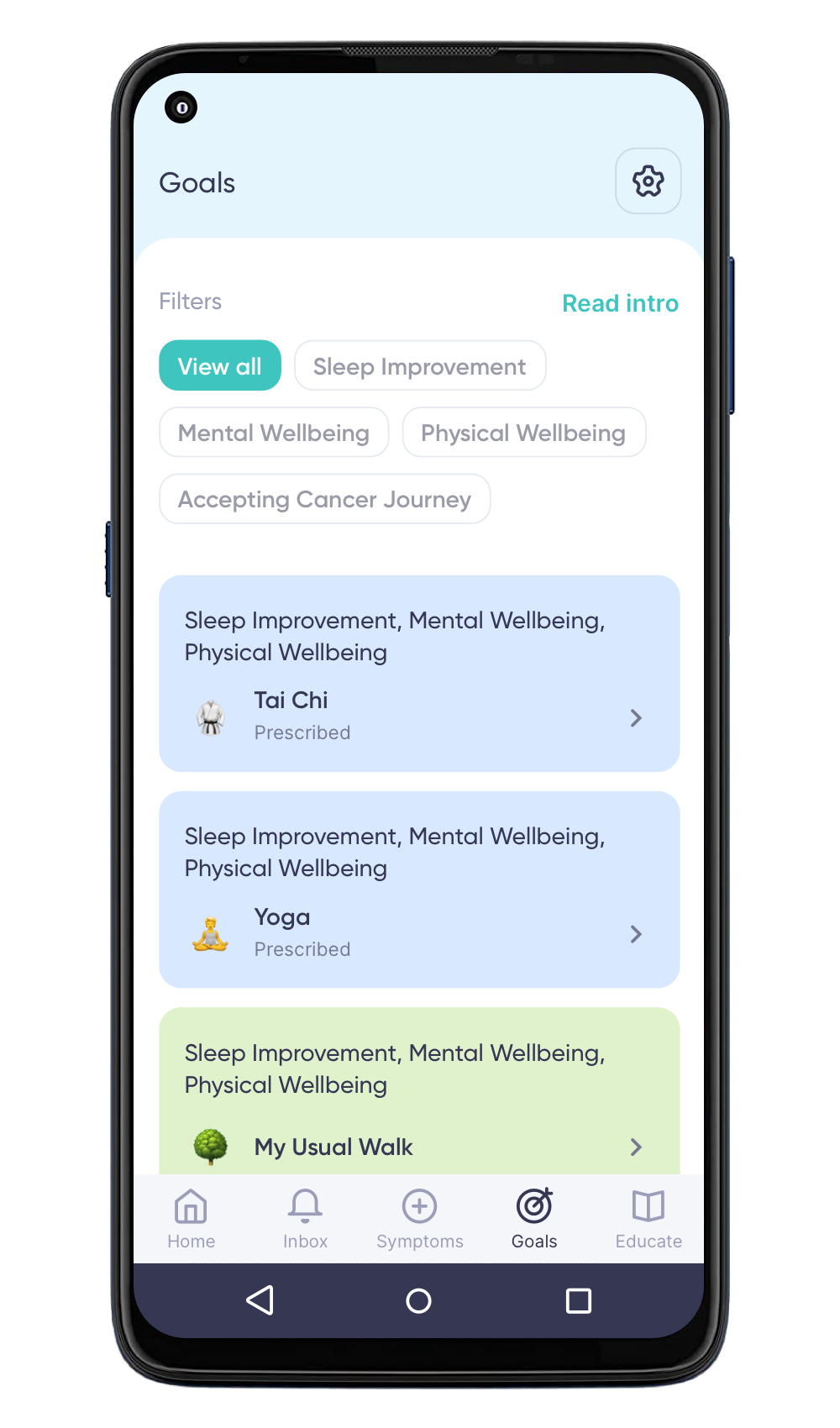}     \label{mpat} }}%
        \subfloat[\centering Goals progress tracking ]{{\includegraphics[width=5.0cm]{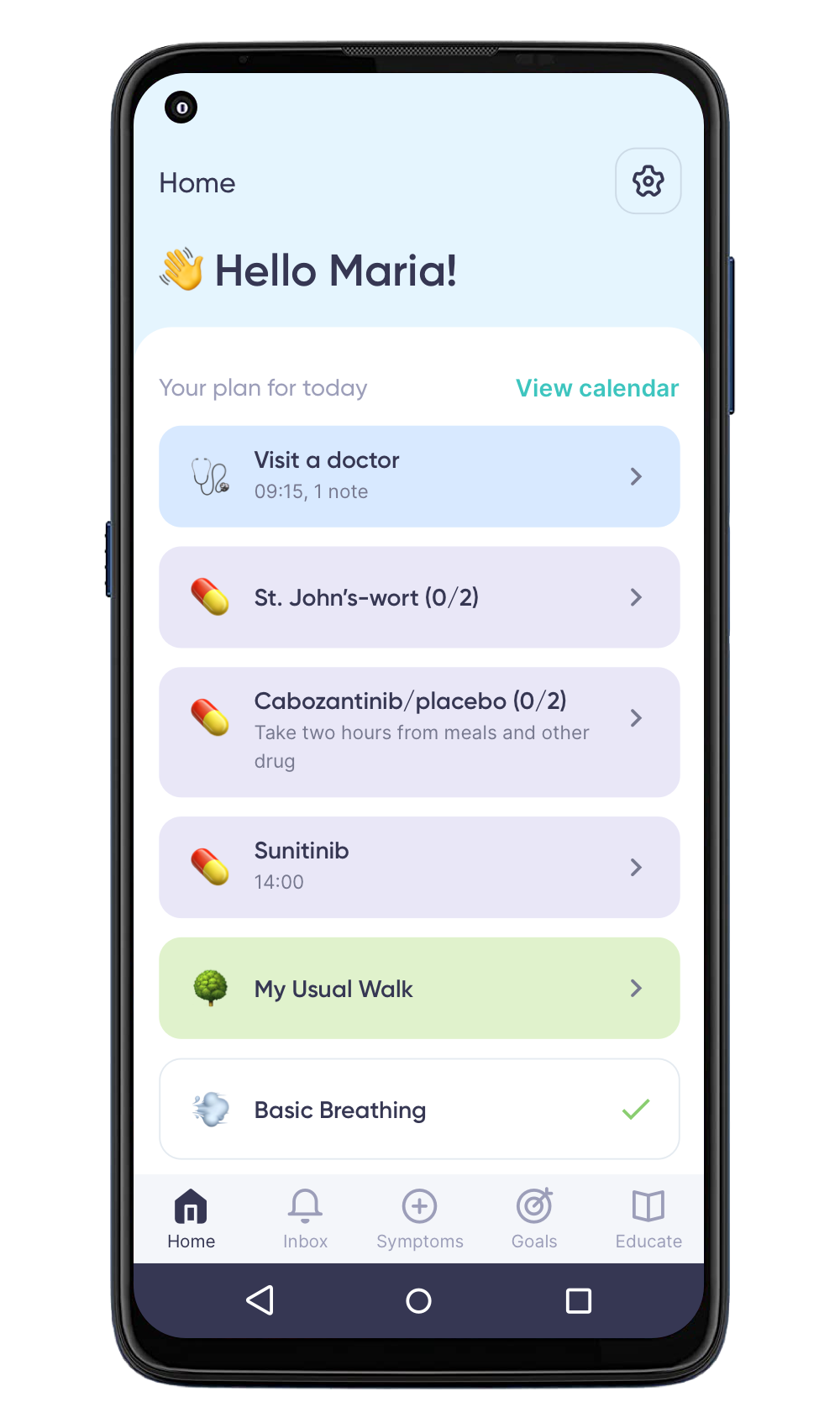}     \label{mpatrev} }}%
    \caption{(a) Mockup of a goal-setting screen in the physicians' app that is used during shared decision-making to set up goals for the patient (agree on a behavioural contract and review behavioural goals); (b) Mockup of the patient app showing BCIs. (c) Mockup of the goal-review screen with feedback on how many of the behaviour goals were achieved compared to the set target. Mock-ups were created by Bitsens UAB, a partner of the CAPABLE Consortium.
}%
    \label{fig:mock_up}%
\end{figure}

Once the patient population (e.g., cancer patients who are fatigued), their context (e.g., at home), clinical goals and some indication of evidence-based BCIs meeting the goals are identified from evidence-based sources, the development team use agile software development methods \cite{lucassen2016improving}, including development of stereotypical persona and user stories, to create a shared understanding of the anticipated user experience with the DHI. Figure \ref{fig:user_story} presents a set of user stories for a patient persona called Maria, and her physician Giulio, that were developed for the CAPABLE project that demonstrate in a high-level way the anticipated user-experience. These user stories form the starting point for app screen mockups (e.g., Figure \ref{fig:mock_up}) and are refined with users' feedback.

\subsubsection{Find intervention protocols and source}
In further iterations of the Ideate step, the informaticians on the team searched for specific existing sources, i.e., implementations for the different interventions, in the form of narrative instructions for deep breathing \cite{WEbMD} or videos demonstrating physical exercise such as Tai Chi \cite{amorim2019integrating}. These were accepted by the clinicians and patients of the multidisciplinary app-design team.

\subsubsection{Select Behaviour Change Techniques}

Abraham and Michie \cite{abraham2008taxonomy} created a taxonomy of behaviour change techniques, which initially included 26 distinct BCTs and later was extended to 93 BCTs \cite{michie2013behavior}.  In BCI content, we bundle together multiple BCTs to create modules reusable in different BCIs (See Table \ref{tab:bci-def} "BCI\_Content"). For the sake of simplicity, we use the original set of 26 BCTs in our examples.

\begin{figure}%
    \centering
    \subfloat[\centering App on-boarding ]{{\includegraphics[width=4.3cm]{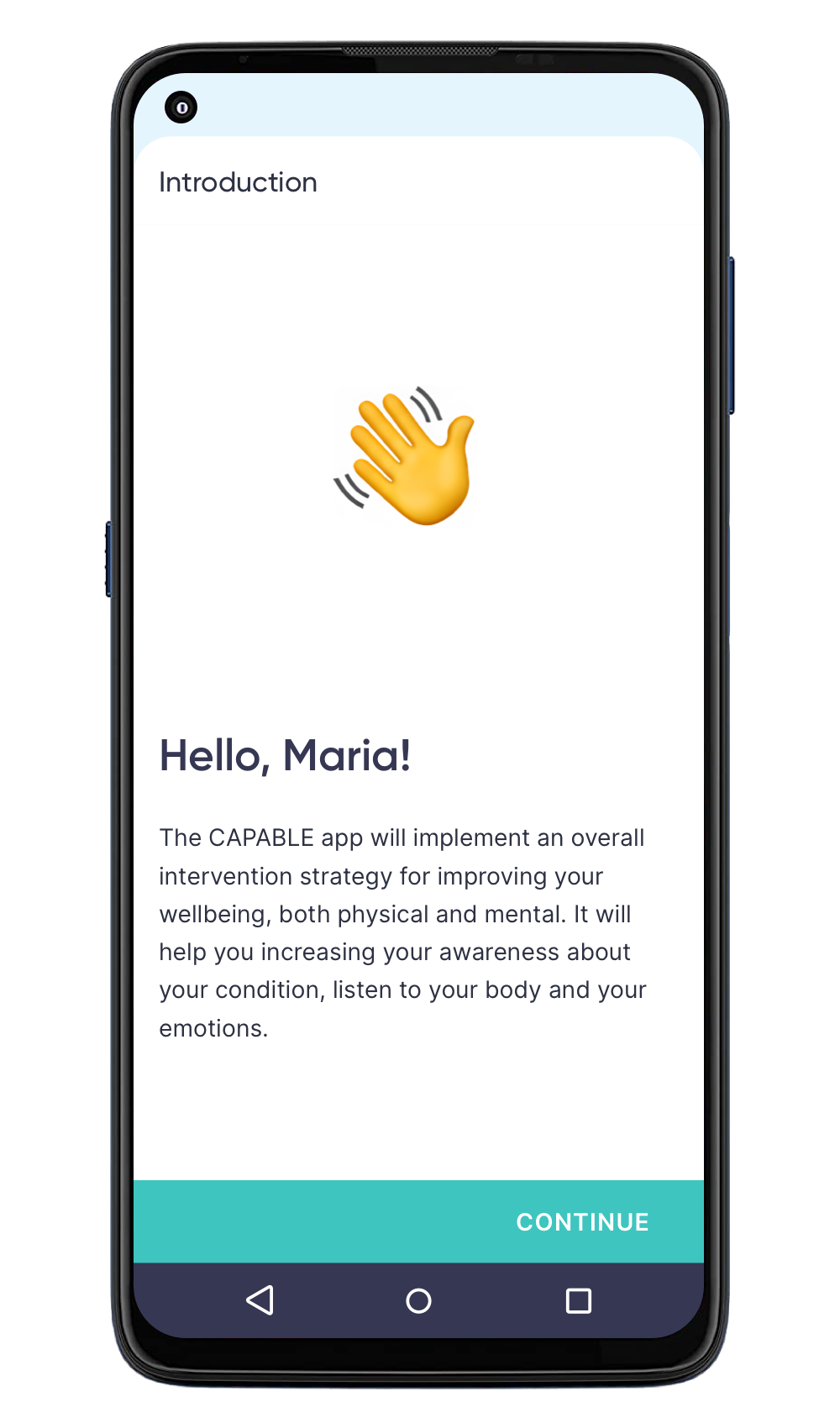}
        \label{edonboarding}}}%
        \subfloat[\centering Tai Chi Information ]{{\includegraphics[width=4.3cm]{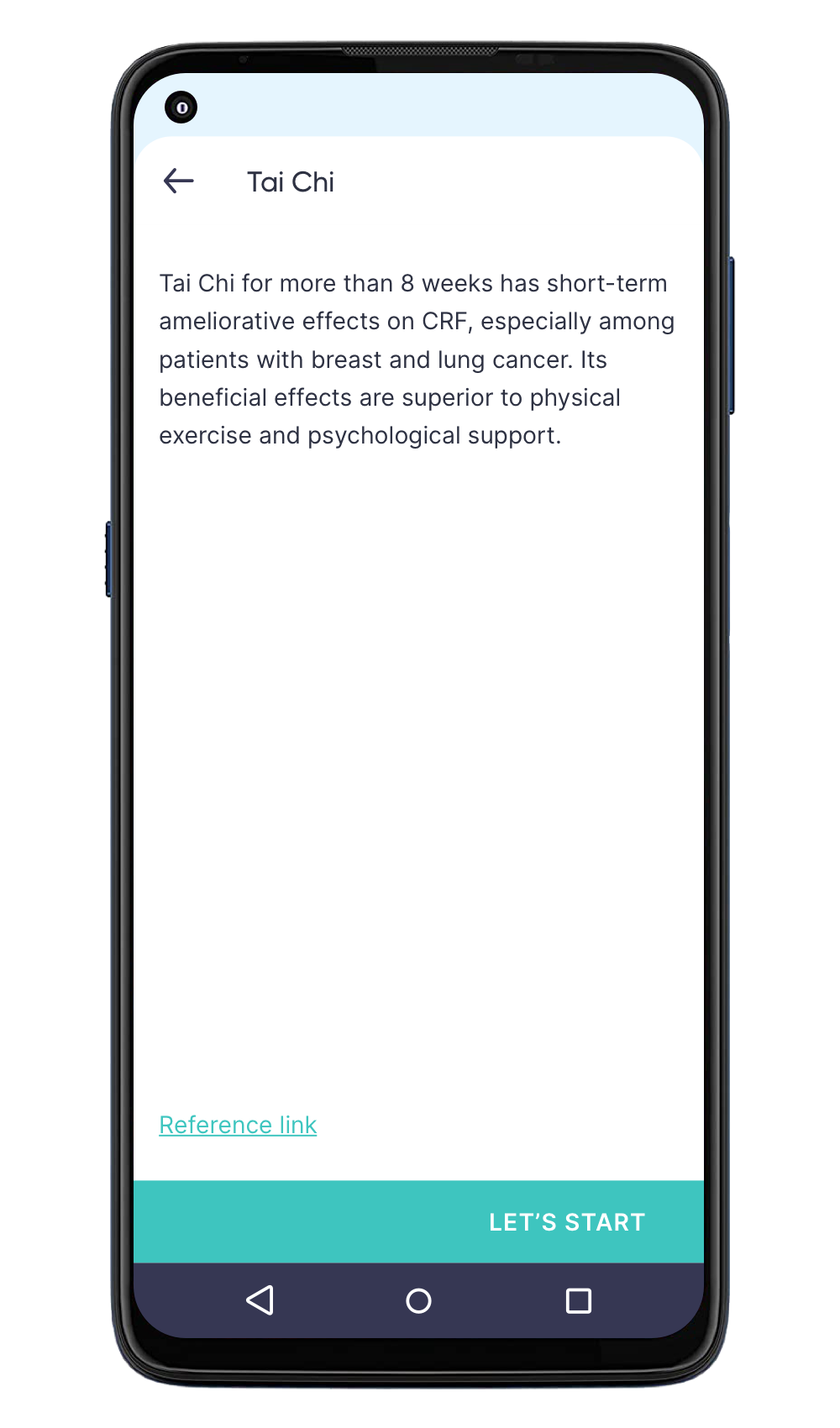}
        \label{tinfo} }}%
        \subfloat[\centering Tai Chi Lessons ]{{\includegraphics[width=4.3cm]{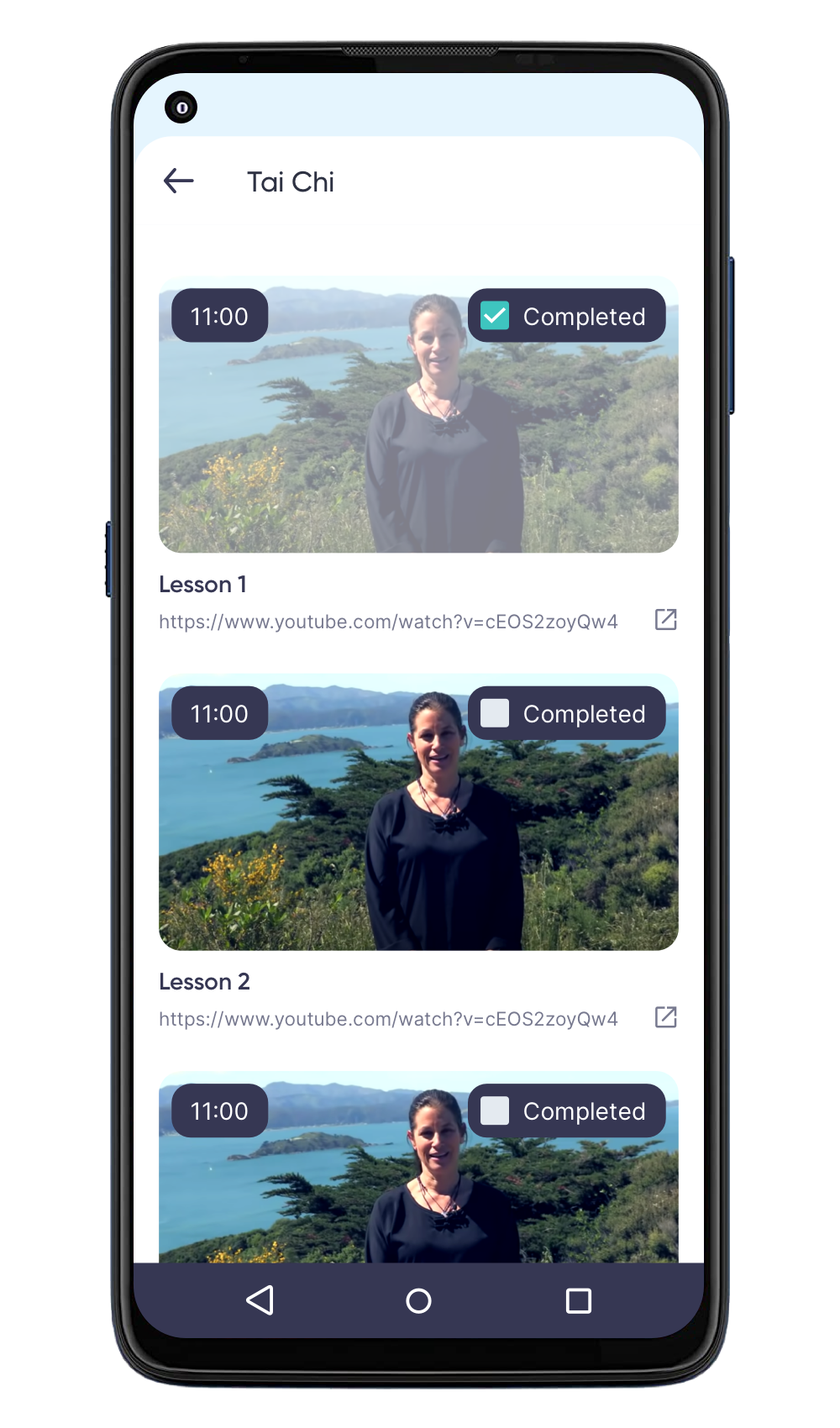}
        \label{tlesson} }}%
    \caption{Mockup screens for education content at the application and individual BCI level. Mock-ups were  created by Bitsens UAB, a partner of the CAPABLE Consortium}%
    \label{fig:edu}%
\end{figure}

Note that some BCT bundles could be applied on clinical goal level as well as at single BCI level.  For example, during onboarding educational materials providing information about the clinical goal and the overall approach of addressing it via behaviour change is explained along with illustrative examples of some of the BCIs. This content fits with BCTs of \textit{providing information regarding behaviour-health link} as well as \textit{information on consequences} (Education\_Content). In the BCIO, BCTs are defined for BCIs and not for the entire BCI Scenario. However, we found it as an essential part of the BCI Scenario Plan to introduce such information about the overall DHI approach (See Figure \ref{edonboarding}) in addition to providing such information at the level of the BCIs (See Figure \ref{tinfo}) of the BCI Scenario.

Although development of the BCTs for a given BCI can be done at any order, we present the BCTs according to the order in which a user is usually exposed to the BCTs over their use of a BCI delivered via the DHI: providing information and instruction (Education\_Content), goal setting and habit planning (Habit\_Planning\_Content), behaviour demonstration/instructions and self-monitoring (Behaviour\_Lesson\_Content e.g. Tai Chi), review of behaviour goals and progress\_tracking (Review\_Content), prompting, and providing encouragement and rewards (Notification\_Content). Most of the app modules supporting these BCTs are directly suitable for reuse in different BCIs with little adaptation. 

Table \ref{tab:bci-def}  provides BCIO examples for the Fatigue Improvement Scenario of the Context "fatigued cancer patients at home". For the goal of less fatigue, several BCIs are developed, each around one Outcome Behaviour. For Tai Chi BCI, we show the details of different BCI\_Content objects, bundling one or more BC Techniques that support habit formation of the Outcome behaviour. The BCI\_Content objects shown include (a) habit planning for Tai Chi (bundling 4 BCTs of intention formation, goal setting, learning to identify a prompt, and time management) and (b) Tai Chi Lesson (bundling the BCTs for demonstrating the behaviour and prompting self-monitoring)

Figure \ref{tlesson} shows Tai\_Chi\_Lesson\_Content mockup with video demonstrations of the activity and ticks to monitor and reward lesson completion. Figure \ref{mpatrev} shows Review\_Content, prompting review of the weekly goals and providing feedback on progress toward the set goal.

\subsection{Customise and Personalise}
Customisation (or customisability) refers to a creation of predefined set of options during the design step, whereas personalisation is a process of matching patients to the options best suited for them. Personalisation is one of the most commonly used techniques in mobile health interventions \cite{dugas2020unpacking} and it plays important role in influencing their effectiveness \cite{gosetto2020personalization}.

Tong et al.\cite{tong2021personalized} conducted a systematic review of personalized mobile BCIs and highlighted that personalisation might be applied to: BCI\_Content (e.g., demonstration video),  BCI\_Mode\_of\_delivery (e.g., voice message, game, wearable), BCI\_Dose (e.g., number of daily notifications), BCI\_Schedule\_of\_delivery.   Automatic, data-driven personalisation depends also on the type and source of the collected data, the frequency of data collection, and the personalisation algorithm. These should be also already considered at the customization option-generation step.
% Personalisation or tailoring refers to  the  process  of adapting intervention to the specific characteristics of an  individual.

\subsubsection{Define set of customisation options}

To maximize the impact of education information on patients’ outcome behaviour, it is important that patients perceive the information to be personally relevant. Ghalibaf \etal conducted a systematic review of computer-based health information tailoring and identified six dimensions according to which patients could be characterized, these are: 1) socio-demographic (e.g., age, level of education), 2) medical history(e.g., comorbidities), 3) health state (e.g., disease  severity), 4) psycho-behavioural determinants (e.g., attitude, self-efficacy) 5) knowledge level, 6) history of interactions (e.g., visited pages) \cite{ghalibaf2019comprehensive}. In practice, majority of studies used three or fewer dimensions for user profiling with socio-demographic and psycho-behavioural features being the most popular.

The choice of the user categorisation dimensions depends on the BCI Content. Some parts are \textit{static} ( ie, selected once prior to interventions commencement), other are \textit{dynamic} (ie, change depending on the user interaction with the application). For example, Michalowski \etal \cite{michalowski2021health} developed precompiled educational plans depending on patients education level and their preferred learning style (e.g., visual or audio). To further enhance patients’ understanding of their care the authors included Question and Answer (Q\&A) components in the education plan. The tailoring of Q\&A difficulty level was based on Blooms taxonomy of educational objectives \cite{bloom1956taxonomy} and depended on the patient’s learning progress captured in the response log.

Another interesting mode of delivery supporting long-term engagement with BCI, and worth considering during the customisation step, is gamification \cite{cugelman2013gamification}. DHI gamification frequently relies on social cognitive theory (SCT), which states that people learn new behaviours through observation of others \cite{bandura1977social} and the behavioural change goal is realized through self-observation, self-evaluation, self-reaction and self-efficacy. The BCI Engagement could be created to align with SCT and include gamification elements, e.g. the notification template for Tai Chi Notification Content might read as follows: \textit{”The majority of patients felt relief of fatigue symptoms after one month of daily Tai Chi practice. You have completed X Tai Chi lessons this week (Y\% of your weekly target and Z lesson less/more than patients similar to you.”} 
A potential problem with this template is that people vary in response to social comparison \cite{lisowska2021personality}; some get motivated others may get discouraged \cite{schmidt2019gamification}.

Tondello \etal created the Hexad Player Type Framework \cite{tondello2016gamification}, which  identifies the following six player types: Disruptor, Free Spirit, Achiever, Player, Socialiser, Philanthropist.  Each player type is associated with a set of game elements that motivates them. For example Socialiser will get motivated by social competition, or team activities, whereas Achiever would prefer challenges and progression through levels. Profiling patients according to their player type could help to match them to adequate notification and weekly goal review templates and facilitate their adherence to BCIs.

Finally, DHI should be customised for different goals. Goal setting is associated with personalisation, but we should be ready up-front for specifying different goals. Baretta \etal \cite{baretta2019implementation} reviewed multiple apps directed towards encouraging physical activity in the context of goal setting theory. In particular, they evaluated components of goal setting: specificity, difficulty, time-frame, action planning, evaluation and reevaluation. The authors found that the majority of the apps enabled specific goal setting and many considered time-frame but few allowed the tailoring of goal difficulty to the users’ ability level and non considered re-evaluating the goals based on individuals achievements. 
In CAPABLE the goal setting is spread across multiple BCT bundles. In Habit Planning Content, patients set the time in which they want to perform the target behaviour. In Behaviour Lesson Content, the user selects the length of the lesson and in Review Content, the user modifies their weekly goals. Note that the latter two could be also set automatically by an algorithm that selects and enumerates the customisable templates for varying goals and difficulty, which are created during the customisation step.

\subsubsection{Develop personalisation methods}

The goal of the personalisation algorithm is to match the patients to one of available customization options and maximise the probability that they perform the target behaviour. According to Fogg’s Behaviour Model (FBM) \cite{fogg2009behavior}, three factors impact behaviour completion: motivation, ability and trigger. Tailoring of Notification Content and Education Content may increase patients’ motivation, Behaviour Lesson Dose, the user’s ability to perform target behaviour, and schedule of prompt delivery the user’s responsiveness to the notification.

The personalisation algorithms are either knowledge-based and data-based. Sadasivam \etal\cite{sadasivam2016impact} showed that using Bayesian probabilistic matrix factorization, trained on previous message ratings, yields message tailoring that is more effective at influencing users to quit smoking than a rule-based method (Notification Content). Manuvinakurike \etal \cite{manuvinakurike2014automated} utilized an adaptive boosting model to select the best suited health behaviour-change story presented to a user, given their response to questionnaires measuring demographics, self-efficacy, decisional balance (i.e. balance between perceived positive and negative consequences of selecting a new behaviour), and the stage of change, according to the transtheoretical model (TTM) \cite{prochaska2015transtheoretical}.  TTM identifies five stages of change: precontemplation, contemplation, preparation, action and maintenance. Individuals are assumed to move through this stages during the process of behaviour change. The model resulting from considering these data was used for personalized story selection (Education Content) and had a significant impact on users’ self-efficacy when compared to randomly assigned stories.

ML models have also been utilized for tailoring of the notification timing (BCI Schedule of delivery). Morris \etal \cite{morrison2017effect} used information from GPS, accelerometer and time of the day to train a Naive Bayes classifier to predict if the participant will respond to the notification. When ML-powered push-notifications were compared with daily notification within a predefined time frame, there was no difference in user responsiveness. Nevertheless, other ML approaches that aimed at maximization of responsiveness incorporated richer information, such as mobile application usage \cite{pielot2017beyond} or user personality traits \cite{kunzler2019exploring} into model training and reported good performance in responsiveness prediction on retrospective data. 

 We considered using the patient’s internal context such as stress \cite{lisowska2021catching} and cognitive load \cite{lisowska2021good}, derived from data captured by the watch, alongside external context information such as location and time of the day for learning the best time to intervene. Given that prior to the intervention we had no data about patients’ responsiveness in varying contexts, we conducted an experiment on simulated data  \cite{lisowska2021personalized} to evaluate different ML approaches to notification schedule personalisation and found adaptive supervised learning approaches to be more effective than reinforcement learning methods.

Rule-based approaches might be well suited for automatic BCI Goal Setting. For example, when a patient has short and inconsistent sleep, a rule may suggest a BCI supporting sleep quality improvement. The BCI Dose might be also implemented as a set of rules, starting from the recommended value for the target population (eg, Tai Chi 3 times per week) and adjusted according to the number of times the patient performed the activity during the last week (eg, reduced to 2 times a week if the goal was not met) to fit the patient’s ability.

\section{Evaluation}
We evaluated the applicability of our workflow and checklist by considering multiple clinical goals/BCI Scenarios, BCIs (Capsules), and BCT Bundles. We found that our methods were appropriate for all of these scenarios, BCIs and BCT bundles.  
In Table \ref{tab:ev} we summarise goals and interventions for which we defined content as part of CAPABLE app. 

\begin{table}[ht]
\centering
\caption{Goals and content defined in CAPABLE app}
\label{tab:ev}
\begin{tabular}{llc}
\hline
 & Names & \multicolumn{1}{l}{Total Number} \\ \hline
\begin{tabular}[c]{@{}l@{}}Goals\\ BCI Scenarios\end{tabular} &
  \begin{tabular}[c]{@{}l@{}}Fatigue  Reduction \\ Sleep Improvment\\ Stress/Anxiety Reduction\\ Mood Improvment\\ Physical Activity Increase\end{tabular} &
  5 \\ \hline
\begin{tabular}[c]{@{}l@{}}BCIs (Capsules)\end{tabular} &
  \begin{tabular}[c]{@{}l@{}}Deep Brathing, Imagery Training, \\  Tai Chi, Yoga,  Garden Bowl,  \\Gratitude Journal, Photo Collage\end{tabular} &
  7 \\ \hline
BCT Bundles &
  \begin{tabular}[c]{@{}l@{}}Education\_Content \\ Habit\_Planning\_Content \\ Behaviour\_Lesson\_Content\\ Review\_Content\\ Notification\_Content\end{tabular} &
  5 \\ \hline
\end{tabular}
\end{table}

\section{Discussion}
We proposed SATO, a DHI design workflow aligned with the BCIO and the IDEAS framework, which we evaluated as being comprehensive for designing several BCI Scenarios for a CAPABLE System app for cancer patients, as part of the CAPABLE project. To our knowledge, this is a first DHI guide which considers multiple BCIs and builds a Goal hierarchy with defined evaluation metrics at each level. We also provided examples on utilising behaviour change theories, such as SCT and FBM, when considering customisation and personalisation of the intervention for maximum engagement and adherence.

Although we presented the design steps consecutively, in practice they are iterative (See Figure \ref{fig:workflow}). For example, at the point of defining the target population and setting, we started creating the user stories for the DHI and later refined them for each BCI Scenario and BCI. Similarly, the lead measures were changed after testing the real capabilities of the selected smartwatches. Moreover, the knowledge engineers on our team suggested to incorporate a wide range of BCTs; nevertheless some content elements (eg, providing feedback through progress visualisation) were not included in the final app because the psychologists who were part of the multi-stakeholder development team raised concern that the ability of cancer patients to perform target behaviour might actually deteriorate with time due to the toxic effects of the cancer therapy and the course of illness; in such cases, visualizing progress data might negatively impact their emotional well-being. This example highlights the need for iterative redesign with the end users being continuously kept in mind.

The proposed guide focuses on DHI design, therefore the Share step from IDEAS framework is not comprehensively addressed. We have however utilized BCIO and extended it to facilitate knowledge sharing. The study with the cancer patients has not yet commenced, hence we could not provide concrete examples of Evaluation Findings. We will address this limitation and also evaluate the suitability of our chosen lead measures in future work, when conducting the clinical study of the usage of the CAPABLE system by cancer patients.

\section{Conclusions}
We described a process of designing digital behaviour change intervention which incorporates a range of behaviour change techniques and addresses multiple clinical goals. The step-by-step SATO guide that we created extends BCIO to support the scenarios of multiple intertwined and hierarchical goals and therefore could be used for design of any non-pharmacological digital health intervention.  We aimed to keep the process simple and provide concrete examples of: technology-independent system captured lead metrics, application modules bundling several behaviour change techniques, and customisation templates based on behaviour change theories which could be readily reused in other DHIs. 

\section*{Acknowledgments}
The CAPABLE project has received funding from the European Union's Horizon 2020 research and innovation programme under grant agreement No 875052. This work has been also partly supported by the EU H2020 grant Sano No. 857533 and the IRAP Plus programme of the Foundation for Polish Science. 

The authors also wish to thank Vitali Gisko and Valentina Ganicheva from Bitsens UAB for provision of the app mock-ups.

%  \cite{kaissis2020secure}

 \bibliographystyle{vancouver} 
 \bibliography{dbci}

\end{document}